\documentclass[amsmath, amsfonts, superscriptaddress, twocolumn,prb]{revtex4}
\usepackage{graphicx}
\usepackage{epsfig}
\usepackage{bm}
\usepackage{dcolumn}
\usepackage{amsmath}
\usepackage{amssymb}
\usepackage{color}
\usepackage{stackrel}
\usepackage{accents}
\usepackage{latexsym}
\usepackage{verbatim}
\usepackage{hyperref}
\usepackage{bbm}

\newcommand{\beg}{\begin{equation}}
\newcommand{\en}{\end{equation}}

\newcommand{\bp}{\mathbf p}

\newcommand{\bk}{\mathbf k}
\newcommand{\br}{\mathbf r}

\newcommand{\eps}{\epsilon}

\newcommand{\up}{\uparrow}
\newcommand{\dn}{\downarrow}
\newcommand{\dg}{^\dagger}

\begin{document}

\title{Interacting fermions in narrow-gap semiconductors with band inversion}

\author{Suman Raj Panday and Maxim Dzero}
\affiliation{Department of Physics, Kent State University, Kent, Ohio 44242, USA}

\begin{abstract}
Highly unconventional behavior of the thermodynamic response functions has been experimentally observed in a narrow gap semiconductor samarium hexaboride. Motivated by these observations, we use renormalization group technique to investigate many-body instabilities in the $f$-orbital narrow gap semiconductors with band inversion in the limit of weak coupling. By projecting out the double occupancy of the $f$-states we formulate a low-energy theory describing the interacting particles in two hybridized electron- and hole-like bands. The interactions are assumed to be weak and short-ranged. We take into account the difference between the effective masses of the quasiparticles in each band and find that there is only one stable fixed point corresponding to the excitonic instability with time-reversal symmetry breaking for small enough mismatch between the effective masses.
\end{abstract}


\maketitle

\section{Introduction}

Anomalous behavior of thermodynamic response functions at low temperatures more often than not remains a hallmark of strong interparticle correlations in quantum materials.\cite{Coleman2007} Among many examples of such materials are cerium- and iron-based superconductors which develop superconducting order and exhibit unusual temperature dependence in heat capacity and in London penetration depth correspondingly.\cite{Petrovic2001,Thompson2001,Sarrao2007,Matsuda_Science2012,Matsuda_review2014} These thermodynamic anomalies are likely governed by the system's proximity to the underlying magnetic quantum phase transition which mediates a strong interactions between the constituent quasiparticles. \cite{Sasha, Sachdev,Fernandes2020,Khodas2020}

Correlated insulators, just like superconductors discussed above, may also exhibit anomalous thermodynamic properties which are not necessarily related to the strong exchange interactions between the local magnetic moments. In a most recent example, quantum oscillations in magnetization and a low-temperature upturn in the heat-capacity have been observed in a correlated narrow gap semiconductor samarium hexaboride. \cite{Suchitra2017,SmB6HC1} The experimental interest in this material, which dates back to 1960s,\cite{SmB6,PeterReview} has been recently revived in relation to its unconventional transport properties: below $T^*\simeq 5$K the Ohm's law breaks down so that the bulk develops a gap with respect to the current carrying excitations while only surfaces remain metallic. \cite{Kim2014} 
Theory proposals which would explain such a behavior focus on the possibility of the inversion of even- and odd-parity bands in the high-symmetry points of the Brillouin zone. \cite{Dzero2010,Takimoto2011,Dzero2012,Alexandrov2013,Dzero2016} As a result of the band inversion, the surfaces of the sample remain metallic even though the bulk remains fully insulating. 

The model with an inverted band structure can also be used in the calculation of the quantum oscillations in magnetization. \cite{Knolle2015,Onur2016,FaWang2016,Inti2018,Fu2018,Pal2019HC} A self-consistent theory of the low-temperature upturn in the heat capacity likely demands that one would need to go beyond a non-interacting low-energy theory. Such an attempt was made by Knolle and Cooper, who formulated a low-temperature theory first by projecting out the double occupation on the $f$-orbitals and, then included the interaction terms, which ultimately lead to the formation of excitons and magnetoexcitons. \cite{Knolle2017} It is not \emph{a priori} clear, however, which instability - if any of the two -  would be the leading one. Furthermore, one may consider a scenario in which superconductivity competes with the excitonic-type of instability upon doping these materials with carriers. 

In what follows, we address this problem by formulating the low-energy theory with an effective action which includes the short-ranged interactions allowed by symmetry of the problem. Specifically,  we consider a generic two-band model with a hybridization gap as a starting point. The one band is assumed to be electron-like, while the other one is hole-like. The hybridization depends linearly on momentum, which corresponds to the case of the $d$- and $f$-orbital bands, while the parabolic dispersion relation is assumed for both bands. We will not make any specific assumptions on the position of the chemical potential at the beginning of the renormalization group flow.  Importantly, since within our theory the band parameters as well as hybridization amplitude have been renormalized from their bare values by projecting out the doubly occupancy on $f$-orbitals, so that the effective masses for conduction and valence band quasiparticles are not equal. Consequently, the emerging many-body instabilities can be studied by employing the renormalization group (RG) technique. We find that for an arbitrary ratio between the effective masses of the conduction and valence bands there is a fixed point which corresponds to an instability favoring the formation of magnetic excitons. 

This paper has been structured as follows. In the next Section we provide the details on the model, discuss the relevant approximations and write down the low-energy theory with the two-particle interactions included. In Section III we analyze the low-energy theory using the renormalization group approach and derive the renormalization group flow equations for the corresponding coupling constants in both particle-hole and particle-particle channels. Section IV is devoted to the discussion of the results and conclusions.  

\section{Model}
When discussing the materials with partially filled $f$-orbitals, the Anderson lattice model is usually the starting point. Since the contact interactions between the $f$-electrons is the largest energy scale in the problem, in order to formulate the low-energy theory one usually projects out the doubly occupied states on $f$-orbitals.\cite{Coleman2007} This procedure leads to the renormalization of the parameters in the Anderson model Hamiltonian. This low-energy model Hamiltonian is our starting point.
\subsection{Single-particle action}
We consider the following single particle Hamiltonian:
\beg\label{Eq1}
\hat{H}_0=\sum\limits_{\bk}{\Psi}_{a}\dg(\bk)\left[
\begin{matrix}
\epsilon_c(\bk)\hat{\tau}_0 & \hat{\Phi}_\bk \\
\hat{\Phi}_\bk\dg & \epsilon_f(\bk)\hat{\tau}_0
\end{matrix}
\right]_{ab}\Psi_{b}(\bk),
\en
where ${\Psi}_\bk\dg=(c_{\bk\up}\dg, ~c_{\bk\dn}\dg, ~f_{\bk\up}\dg, ~f_{\bk\dn}\dg)$, $\bk=(k_x,k_y,k_z)$ is the momentum,  $\hat{\tau}_0$ is a 2$\times$2 unit matrix, $\hat{\Phi}_\bk$ is a 2$\times$2 hybridization matrix to be specified below and $c,f$ are fermionic annihilation operators for the conduction and valence bands. It is convenient to write the single particle dispersion relation as \cite{Fu2018}
\beg
{\epsilon_c(\bk)=\frac{k^2}{2m_c}+\frac{E_g}{2}+\mu_0, ~\epsilon_f(\bk)=\frac{k^2}{2m_f}+\frac{E_g}{2}-\mu_0},
\en
where $E_g<0$ is the energy gap, $\mu_0$ is the energy shift and it is implicitly assumed that $m_f>m_c$. 
It will be convenient to introduce
${m_{\pm}}^{-1}={m_c}^{-1}\pm{m_f}^{-1}$ and $k_F^2=-2m_{+}E_g$. If we now set $\mu_0=-k_F^2/4m_{-}$ it follows
\beg\label{eBands}
\begin{split}
\epsilon_c(\bk)&=\frac{k^2-k_F^2}{4m_+}+\frac{k^2-k_F^2}{4m_{-}}\equiv\xi_\bk+\eps_\bk, \\
\epsilon_f(\bk)&=\frac{k^2-k_F^2}{4m_{-}}-\frac{k^2-k_F^2}{4m_+}\equiv\xi_\bk-\eps_\bk.
\end{split}
\en

The specific form and momentum dependence of matrices entering into (\ref{Eq1}) is determined by the type of the hybridizing orbitals. 
Here we consider the fairly standard form corresponding to the hybridization between even- and odd-parity orbitals with angular momentum transfer of $\Delta l=1$:
\beg\label{Phik}
\hat{\Phi}_\bk=V\left(\begin{matrix}
k_z & k_x-ik_y \\
k_x+ik_y & -k_z
\end{matrix}
\right).
\en

With the help of the Dirac matrices, listed in the Appendix A, the single-particle part of the action reads
\beg\label{L0}
S_0=\int_x{\Psi}\dg(x)\left(\frac{\partial}{\partial \tau}-\mu+\xi_{\hat{\bk}}\mathbbm{1}_4+\sum\limits_{a=0}^3\Sigma^ad_{\hat{\bk}}^a\right)\Psi(x),
\en
where $\mu$ is the chemical potential, $x=(\br,\tau)$ and 
\beg\label{SigmasDs}
{\begin{split}
&\Sigma^0=\gamma_0, \quad \Sigma^1=\gamma_0\gamma_1, \quad \Sigma^2=\gamma_0\gamma_2, 
\quad \Sigma^3=\gamma_0\gamma_3, \\
&d_\bk^0=\eps_\bk, \quad d_\bk^1=Vk_x, \quad 
d_\bk^2=Vk_y, \quad d_\bk^3=Vk_z.
\end{split}}
\en
Clearly, when $m_c=m_f$, the term proportional to $\mathbbm{1}_4$ is zero. 

\subsection{Interactions}
The most general form of the Lagrangian density describing weak repulsive interactions is \cite{Bitan0,Bitan}
\beg\label{Lint}\nonumber
{
\begin{split}
&{\cal L}_{\textrm{int}}=\tilde{g}_1\left({\Psi}\dg\Psi\right)^2+\tilde{g}_2\left({\Psi}\dg \tau_1\sigma_0\Psi\right)^2+\tilde{g}_3\left({\Psi}\dg \tau_2\sigma_0\Psi\right)^2\\&+\tilde{g}_4 \left({\Psi}\dg \tau_3\sigma_0\Psi\right)^2
+\tilde{g}_5\left[\left({\Psi}\dg \tau_2\sigma_1\Psi\right)^2+\left({\Psi}\dg \tau_2\sigma_2\Psi\right)^2\right.\\&\left.+\left({\Psi}\dg \tau_2\sigma_3\Psi\right)^2\right]+
\tilde{g}_6\left[\left({\Psi}\dg \tau_3\sigma_1\Psi\right)^2+\left({\Psi}\dg \tau_3\sigma_2\Psi\right)^2\right.\\&\left.+\left({\Psi}\dg \tau_3\sigma_3\Psi\right)^2\right]
+\tilde{g}_7\left[\left({\Psi}\dg \tau_0\sigma_1\Psi\right)^2+\left({\Psi}\dg \tau_0\sigma_2\Psi\right)^2\right.\\&\left.+\left({\Psi}\dg \tau_0\sigma_3\Psi\right)^2\right]+
\tilde{g}_8\left[\left({\Psi}\dg \tau_1\sigma_1\Psi\right)^2+\left({\Psi}\dg \tau_1\sigma_2\Psi\right)^2\right.\\&\left.+\left({\Psi}\dg \tau_1\sigma_3\Psi\right)^2\right].
\end{split}}
\en
Here ${\vec \tau}$ are Pauli matrices act in the band space, while ${\vec \sigma}$ are Pauli matrices in spin space.
Since ${\cal L}=T-U$, the generic behavior corresponds to the case when all coupling constants $\tilde{g}_i$ are negative, i.e. all interactions are assumed to be
repulsive from the outset.
Furthermore, I introduce basis matrices ${\vec \Gamma}$ according to 
\beg\label{Gammaj}
\begin{split}
&\Gamma^1=\mathbbm{1}_4, ~\Gamma^2=\tau_0\sigma_1, ~ \Gamma^3=\tau_0\sigma_2, ~ \Gamma^4=\tau_0\sigma_3, \\
& \Gamma^5=\tau_1\sigma_0, ~ \Gamma^6=\tau_1\sigma_1, ~ \Gamma^7=\tau_1\sigma_2, ~\Gamma^8=\tau_1\sigma_3, \\
&\Gamma^9=\tau_2\sigma_0, ~ \Gamma^{10}=\tau_2\sigma_1 , ~ \Gamma^{11}=\tau_2\sigma_2, ~ \Gamma^{12}=\tau_2\sigma_3, \\
&\Gamma^{13}=\tau_3\sigma_0, ~\Gamma^{14}=\tau_3\sigma_1, ~\Gamma^{15}=\tau_3\sigma_2, ~ \Gamma^{16}=\tau_3\sigma_3.
\end{split}
\en
Importantly, each of these matrices satisfies
\beg\label{basis}
{\left(\Gamma^a\right)\dg=\Gamma^a=\left(\Gamma^a\right)^{-1}.}
\en
Below we will show that not all interaction terms are independent from each other and, as a result, Eq. (\ref{Lint}) can be further simplified.\cite{Bitan0,Bitan,BilayerOskar}
\subsection{Fierz identity}
Thus, we have eight coupling constants, $g_j<0$, However, only four of these matrices (and the corresponding couplings) are independent. 
To prove this, let us employ the following Fierz identity \cite{Bitan0,Bitan,BilayerOskar}
\beg\label{Fierz}
\begin{split}
&\left({\Psi}\dg(x)M\Psi(x)\right)\left({\Psi}\dg(y)M\Psi(y)\right)\\&=-\frac{1}{16}\sum\limits_{ab}\textrm{Tr}\left({M}{\Gamma}^a{M}{\Gamma}^b\right)\left[{\Psi}\dg(x)\Gamma^b\Psi(y)\right]\\&\times\left[{\Psi}\dg(y)\Gamma^a\Psi(x)\right]
\end{split}
\en
along with the relation 
\beg\label{identity}
\delta_{il}\delta_{kj}=\frac{1}{4}\sum\limits_{a=1}^{16}\Gamma_{ik}^a\Gamma_{jl}^a.
\en
Consider now the following vector
\beg\label{VT}\nonumber
\begin{split}
&{\vec V}=\left\{{\left(\overline{\Psi}\Gamma^{1}\Psi\right)^2},
\left(\overline{\Psi}\Gamma^{2}\Psi\right)^2+\left(\overline{\Psi}\Gamma^{3}\Psi\right)^2+\left(\overline{\Psi}\Gamma^{4}\Psi\right)^2,\right.\\&\left.
{\left(\overline{\Psi}\Gamma^{5}\Psi\right)^2},
\left(\overline{\Psi}\Gamma^{6}\Psi\right)^2+\left(\overline{\Psi}\Gamma^{7}\Psi\right)^2+\left(\overline{\Psi}\Gamma^{8}\Psi\right)^2, \right.\\&\left.
{\left(\overline{\Psi}\Gamma^{9}\Psi\right)^2},\left(\overline{\Psi}\Gamma^{10}\Psi\right)^2+\left(\overline{\Psi}\Gamma^{11}\Psi\right)^2+\left(\overline{\Psi}\Gamma^{12}\Psi\right)^2,\right.\\&\left.{\left(\overline{\Psi}\Gamma^{13}\Psi\right)^2},
\left(\overline{\Psi}\Gamma^{14}\Psi\right)^2+\left(\overline{\Psi}\Gamma^{15}\Psi\right)^2+\left(\overline{\Psi}\Gamma^{16}\Psi\right)^2
\right\}.
\end{split}
\en
This choice is matched by the following vector of couplings ${\vec g}=(\tilde{g}_1,\tilde{g}_7,\tilde{g}_2,\tilde{g}_8,\tilde{g}_3,\tilde{g}_5,\tilde{g}_4,\tilde{g}_6)$.
Employing (\ref{Fierz}) along with the definition of vector ${\vec V}$ above, the following system of linear equations 
$\sum_{j=1}^8 {\cal F}_{ij}V_j=0$ obtains with
\beg\label{FV}
{\cal F}=\frac{1}{8}
\left(
\begin{matrix}
5 & 1 & 1 & 1 & 1 & 1 & 1 & 1 \\
3 & 3 & 3 & -1 & 3 & -1 & 3 & -1 \\
1 & 1 & 5 & 1 & -1 & -1 & -1 & -1 \\
3 & -1 & 3 & 3 & -3 & 1 & -3 & 1 \\
1 & 1 & -1 & -1 & 5 & 1 & -1 & -1 \\
3 & -1 & -3 & 1 & 3 & 3 & -3 & 1 \\
1 & 1 & -1 & -1 & -1 & -1 & 5 & 1 \\
3 & -1 & -3 & 1 & -3 & 1 & 3 & 3
\end{matrix}
\right)
\en
The vector of the eigenvalues of this matrix is
${\vec \lambda}=(1,1,1,1,0,0,0,0)$.
Since there are four zero eigenvalues, we have only four independent coupling constants. It will be conveniet to keep the interaction terms with couplings $\tilde{g}_1$, $\tilde{g}_2$, $\tilde{g}_3$ and $\tilde{g}_4$. Lastly, with the help of (\ref{FV}) we can express the remaining interaction terms in terms of the independent ones, which is equivalent to the following change of the coupling constants:
$g_1=\tilde{g}_1-\tilde{g}_5-\tilde{g}_6-2\tilde{g}_7-\tilde{g}_8$, $g_2=\tilde{g}_2+\tilde{g}_5+\tilde{g}_6-\tilde{g}_7-2\tilde{g}_8$,
$g_3=\tilde{g}_3-2\tilde{g}_5+\tilde{g}_6-\tilde{g}_7+\tilde{g}_8$ and $g_4=\tilde{g}_4+\tilde{g}_5-2\tilde{g}_6-\tilde{g}_7+\tilde{g}_8$.
Thus, the interaction part of the Lagrangian density becomes
\beg\label{LintRed}
{
\begin{split}
{\cal L}_{\textrm{int}}(\br,\tau)&={g}_1 \left({\Psi}\dg\Psi\right)^2+{g}_2 \left({\Psi}\dg \tau_1\sigma_0\Psi\right)^2\\&+{g}_3\left({\Psi}\dg \tau_2\sigma_0\Psi\right)^2+{g}_4\left({\Psi}\dg \tau_3\sigma_0\Psi\right)^2.
\end{split}}
\en
Note, that even though $\tilde{g}_j<0$, the renormalized coupling constants $g_i$ can be either positive or negative. 

\section{Renormalization group analysis}
\subsection{Scaling at the tree level}
Each fermionic field is separated into slow ($k<\Lambda/s$) and fast ($\Lambda/s<k<\Lambda$) mode, $\Psi=\Psi_<+\Psi_>$. At the tree level, we need to integrate out the fermions
within the shell of momenta $\Lambda/s<|\bk|<\Lambda$. I have
\beg\label{L0s}
\begin{split}
S_{0<}&=\int\limits_0^\beta d\tau\int_{|\bk|<\Lambda/s}\frac{d^3\bk}{(2\pi)^3}{\Psi}_{<}\dg(\bk,\tau)\left(\frac{\partial}{\partial \tau}-\mu\right.\\&\left.+\xi_\bk\mathbbm{1}_4+\sum\limits_{a}\Sigma^ad_\bk^a\right)\Psi_{<}(\bk,\tau).
\end{split}
\en
Let us rescale momentum back to its initial region $k'\leq\Lambda$ with $k=k'/s$ and $\tau=s^2\tau'$ and replace the fermionic fields
accordingly to keep the action invariant:
\beg\label{S0lessS}
{\Psi}(\bk',\tau')=\frac{1}{\zeta}\Psi_{<}(\bk'/s,s^2\tau').
\en 
It follows
\beg\label{L02}
\begin{split}
S_{0}&=\frac{\zeta^2}{s^3}\int\limits_0^{\beta/s^2} d\tau\int_{|\bk|<\Lambda}\frac{d^3\bk}{(2\pi)^3}{\Psi}\dg(\bk,\tau)\left(\frac{\partial}{\partial \tau}-s^2\mu\right.\\&\left.+\Sigma^0d_\bk^0+s\sum\limits_{a=1}^{3}\Sigma^ad_\bk^a\right)\Psi(\bk,\tau).
\end{split}
\en
Thus, the action remains invariant under the following scale transformation ($s=e^t$):
$T'=s^2T$,  $\mu'=s^2\mu$, $V'=sV$, $m_{\pm}'=s^2m_\pm$,  
$k_F'=sk_F$, $\zeta=s^{3/2}$, 
where the last expression ensures that the action will remain invariant 
and $T$ is the temperature. Clearly, with respect to tree-level perturbations, hybridization coupling $V$ is a
relevant variable under the renormalization group flow. However, hybridization grows slower than the chemical potential. 

\subsection{RG equations: particle-hole channel}
We now proceed with expanding the action in the powers of the interaction up to the second order in powers of $g_j$'s and integrating out the 'fast' modes. 
The effective action in terms of the 'slow' modes is
\beg\label{Basic}
\begin{split}
&\left\langle e^{-S[\Psi]}\right\rangle_{0>}=e^{-S_0[\Psi_<]-S_{\textrm{int}}[\Psi_<]}\langle e^{-S_{\textrm{int}}[\Psi_<,\Psi_>]}\rangle_{0>}\\&= e^{-S_0[\Psi_<]-\delta S[\Psi_<]},
\end{split}
\en
where the $\langle...\rangle_0$ denotes the averaging over the gaussian action and
the interaction part of the action $S_{\textrm{int}}$ is determined by the Lagrangian density (\ref{LintRed}).

We continue with the computation of the average over the fast modes (\ref{Basic}) using the cumulant expansion. Integrating out the fast modes in the momentum shell $k\in[\Lambda/s,\Lambda]$ and rescaling the resulting correction to the effective action using (\ref{S0lessS}) one may find the corrections to the coupling constants. The details of the calculation are 
presented in Appendix B. The resulting flow equations for the four coupling constants are
\beg\label{FlowEqs}
{\begin{split}
\frac{dg_1}{d\ln s}&=-\frac{m\Lambda}{4\pi^2}\left[2g_1g_4+\eta\left(g_1+g_4\right)^2\right.\\&\left.+\eta\left(g_2-g_3)^2\right)\right], \\ 
\frac{dg_2}{d\ln s}&=\frac{m\Lambda}{2\pi^2}
\left[(1-\eta)g_1g_2+\eta g_1g_3+(1+\eta)g_3g_4\right.\\&\left.-g_2g_3-(2+\eta)g_2g_4-g_2^2\right], \\
\frac{dg_3}{d\ln s}&=\frac{m\Lambda}{2\pi^2}
\left[(1-\eta)g_1g_3+\eta g_1g_2+(1+\eta)g_2g_4\right.\\&\left.-g_2g_3-(2+\eta)g_3g_4-g_3^2\right], \\
\frac{dg_4}{d\ln s}&=-\frac{m\Lambda}{4\pi^2}
\left\{(1+\eta)\left[g_1^2+(g_2-g_3)^2+g_4^2\right]\right.\\&\left.+2\eta g_1g_4\right\}.
\end{split}}
\en
Here we use $m=m_{+}$ for brevity, $\Lambda$ is the ultraviolet cutoff. Note, that the second and third equation are symmetric with respect to $g_2\leftrightarrow g_3$. Lastly, parameter $\eta$ describes the mismatch between the effective masses, Eq. (\ref{C1C2simp}), so that the case of two bands with equal effective masses corresponds to the limit $\eta\to 0$.

To analyze Eq. (\ref{FlowEqs}) it will be convenient to work with the coupling ratios.\cite{BilayerOskar} It will be convenient to choose the following coupling ratios: $v_1=g_1/g_4$, $v_2=g_2/g_4$ and $v_3=g_3/g_4$. The flow equations in terms of these variables are easily derived from (\ref{FlowEqs}), so we will write these compactly as
\beg\label{Poly}
\begin{split}
g_4\frac{dv_1}{dg_4}&={\cal R}_1(\eta;v_1,v_2,v_3)-v_1, \\
g_4\frac{dv_2}{dg_4}&={\cal R}_2(\eta;v_1,v_2,v_3)-v_2, \\
g_4\frac{dv_3}{dg_4}&={\cal R}_2(\eta;v_1,v_3,v_2)-v_3.
\end{split}
\en
Since the second and third equations are symmetric with respect to an interchange of $v_2$ and $v_3$, we can determine the fixed points analytically since in order to satisfy the second and third equation simultaneously, we need to require that $v_2^*=v_3^*=v_\perp^*$. The fixed point for the first equation is given by the roots of the following equation:
\beg\label{EqFP}
\begin{split}
&(v_1^*-1)(v_1^*+1)\left(v_1^*+\frac{\eta}{1+\eta}\right)=0,
\end{split}
\en
while the fixed point for the remaining two equations is either $v_\perp^*=0$ or $v_\perp^*=(1/4)[(1+\eta)(v_1^*+1)^2-2]$.
Thus, independent of the value of the parameter $\eta$, Eq.  (\ref{EqFP}), there are six fixed points. 

\begin{figure}[h]
  \centering
 \includegraphics[width=3.35in]{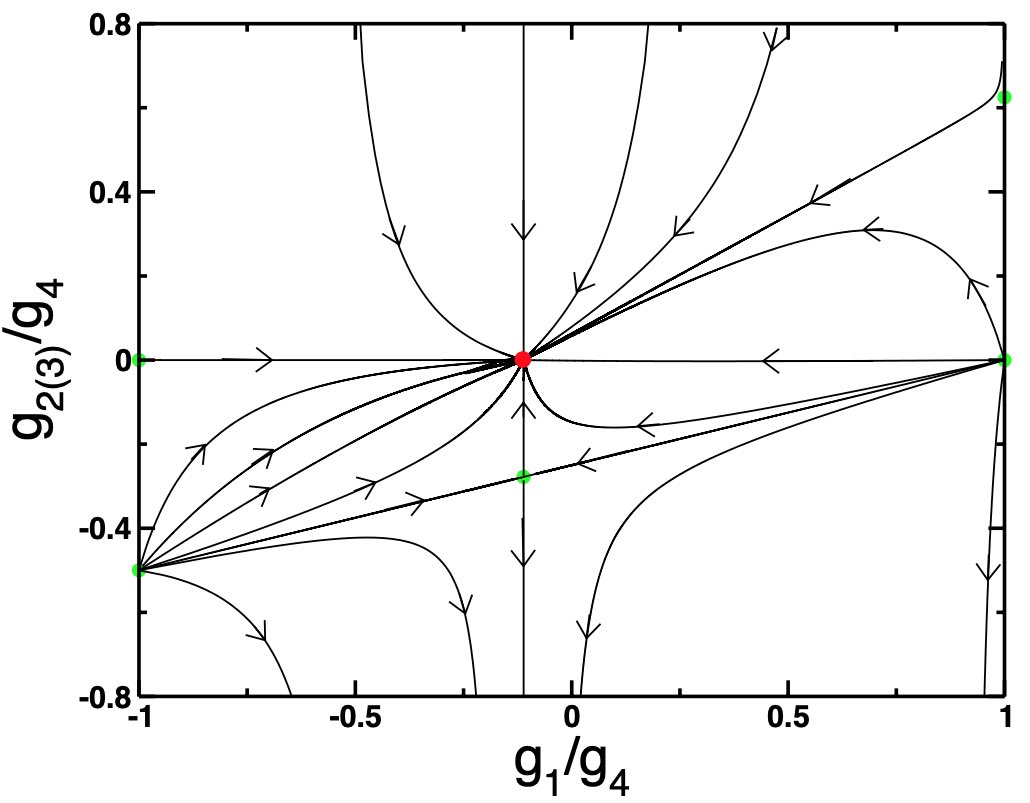}
  \caption{(color online). Renormalization group flow of the coupling constant ratios for the case of small mismatch between the effective masses, $\eta=0.125$.
 We find that there are six fixed points overall in this case. Five fixed points (light green circles) are always unstable.  The remaing one (solid red circle) is stable when $g_4>0$ and becomes unstable when $g_4<0$. Without loss of generality we chose to limit the presentation to a case of $g_2=g_3=g_{2(3)}$ and we also assumed that at the beginning of the RG flow $(m\Lambda/4\pi^2)g_4=0.1$.}
  \label{Fig1}
\end{figure}

Our stability analysis of the flow equations around each fixed point shows that there is only one stable fixed point ("sink"), $\left(-\frac{\eta}{1+\eta},0,0\right)$ when the initial value of the coupling constant $g_4>0$. The remaining five fixed points are all unstable at least in one of the directions in the space
of coupling constants. When the initial value of the coupling $g_4<0$ the flow of the couplings reverses and a stable fixed point becomes a source. 
The resulting RG flow diagram is presented in Fig. \ref{Fig1}.

Given the nature of the materials under discussion, this case is not physically relevant for us. Nevertheless, for completeness, we note that the value of $|g_1/g_4|$ at the stable fixed point equals zero for $\eta=0$ and then it 
increases with an increase in $\eta$, which means that in the absence of mass anisotropy the stable fixed point is a non-interacting one, 
provided $g_4>0$. Lastly, we note that the chemical potential does not effect the flow of the coupling constants as long as $\mu\ll \frac{\Lambda^2}{2m}$ holds.

\subsection{Particle-hole channel susceptibilities} To investigate the leading instability at the stable fixed point in the particle-hole channel we need to analyze the flow of the corresponding susceptibilities. To do that, we modify the action $S\to S+\Delta S$ with\cite{BilayerOskar}
\beg\label{DeltaS}
\begin{split}
\Delta S_{\textrm{p-h}}&=-\chi_{\textrm{ph}}^{(1)}\int d\tau\int\frac{d^3\bk}{(2\pi)^3} {\Psi}\dg(\bk,\tau)\Psi(\bk,\tau)\\&-\sum\limits_{a=2}^{16}\chi_{\textrm{ph}}^{(a)}\int d\tau\int\frac{d^3\bk}{(2\pi)^3} {\Psi}\dg(\bk,\tau)\hat{\Gamma}^a\Psi(\bk,\tau).
\end{split}
\en
Each terms here can be written as a sum of two momentum integrals: one with $k\leq\Lambda/s$ and another with $\Lambda/s\leq k\leq \Lambda$. 
The goal is to determine the change of the susceptibilities under the RG flow by perturbation theory in powers of the coupling constants. The flow equations for the susceptibilities are obtained by expanding the exponent (\ref{Basic}) in powers of $S_{\textrm{int}}[\Psi_<,\Psi_>]+\Delta S[\Psi_>]$ and integrating fermions whose momenta
lie in the outer shell $\Lambda/s\leq k\leq \Lambda$.
Thus the part of the action with the susceptibilities becomes
\beg\label{ReScaleChi2}
\begin{split}
&\Delta S_{\textrm{p-h}}=s^2\int\limits_0^\beta d\tau\int_{|\bk|\leq\Lambda}\frac{d^3\bk}{(2\pi)^3}
\sum\limits_{a=1}^{16}\chi_{\textrm{ph}}^{(a)}\left\{{\Psi}_k\dg
\hat{\Gamma}^a\Psi_k\right.\\&\left.+\sum\limits_{\cal S}g_{\cal S}\Pi_{\Gamma^a{\cal S}}{\Psi}_k\dg
\hat{\Gamma}^a\Psi_k-\sum\limits_{\cal S}g_{\cal S}{\Psi}_k\dg\Upsilon_{\Gamma^a{\cal S}}\Psi_k
\right\},
\end{split}
\en
where $k=(\bk,\tau)$, the summation is performed over the set ${\cal S}=\{\Gamma^1,\Gamma^5,\Gamma^9,\Gamma^{13}\}$ and we use the following notations
\beg\label{PiOMYOM}\nonumber
\begin{split}
\Pi_{{\cal U}{\cal S}}&=\int\limits_{-\infty}^\infty\frac{d\omega_n}{2\pi}\int\limits_{{\Lambda/s}\leq p\leq\Lambda}
\textrm{Tr}\left[G_\bp(i\omega_n){\cal U}G_\bp(i\omega_n){\cal S}\right],\\
\Upsilon_{{\cal U}{\cal S}}&=\int\limits_{-\infty}^\infty\frac{d\omega_n}{2\pi}\int\limits_{{\Lambda/s}\leq p\leq\Lambda}{\cal S}G_\bp(i\omega_n){\cal U}G_\bp(i\omega_n){\cal S}.
\end{split}
\en
After we rescale the momenta and the fermionic fields so that the action takes its original form,  the following equations for the corresponding susceptibilities are 
\beg\label{Flow4chi}
\begin{split}
&\frac{d\ln\chi_{\textrm{ph}}^{(j)}}{d\ln s}=2, ~(1\leq j\leq 4, ~13\leq j\leq16), \\
&\frac{d\ln\chi_{\textrm{ph}}^{(5)}}{d\ln s}=2+\frac{m\Lambda}{2\pi^2}(1-v_1+3v_2+v_3)g_4, \\
&\frac{d\ln\chi_{\textrm{ph}}^{(6,7,8)}}{d\ln s}=2+\frac{m\Lambda}{2\pi^2}(1-v_1-v_2+v_3)g_4, \\
&\frac{d\ln\chi_{\textrm{ph}}^{(9)}}{d\ln s}=2+\frac{m\Lambda}{2\pi^2}(1-v_1-v_2+3v_3)g_4, \\
&\frac{d\ln\chi_{\textrm{ph}}^{(10,11,12)}}{d\ln s}=2+\frac{m\Lambda}{2\pi^2}(1-v_1+v_2-v_3)g_4.
\end{split}
\en
By performing the numerical solution of the flow equations (\ref{FlowEqs}) around the stable fixed point, we find that when both $v_2(s)$ and $v_3(s)$ approach zero from above or when $v_3(s)$ approaches zero from above, while $v_2(s)$ approaches zero from below, the fastest growing susceptibility corresponds to the order parameter
$\phi_s=\langle{\Psi}_\alpha\dg\left((\tau_1\pm i\tau_2){\sigma}_0\right)_{\alpha\beta}\Psi_\beta\rangle$,
which describes the spin-singlet excitonic order. In the opposite case, when both $v_2(s)$ and $v_3(s)$ approach zero from below the fastest growing susceptibility describes the emergence of the magneto-excitonic order with the order parameter 
${\vec \phi}_{t}=\langle{\Psi}_\alpha\dg\left((\tau_{1}\pm i\tau_2){\vec \sigma}\right)_{\alpha\beta}\Psi_\beta\rangle$. Thus, we confirm that the leading instabilities in the particle-hole channel are the instabilities leading to the formation of an excitonic insulator. It remains to be seen whether the superconducting instability may develop faster or not.

\subsection{Particle-particle channel: renormalization group equations} In order to investigate the superconducting instability, the Lagrangian density (\ref{LintRed}) can be recast into the form describing the interactions in the particle-particle channel. This goal can be accomplished with the help of the Fierz identity
\beg\label{ph2pp}
\begin{split}
&\left({\Psi}\dg(x)M\Psi(x)\right)\left({\Psi}\dg(x)M\Psi(x)\right)\\&=\frac{1}{16}\sum\limits_{ab}\textrm{Tr}\left({\Gamma}^aM{\Gamma}^b{M}^T\right)\left({\Psi}\dg(x)\Gamma^a{\Psi}^*(x)\right)\\&\times\left({\Psi}^T(x)\Gamma^b\Psi(x)\right).
\end{split}
\en
The fermionic nature of the fields $\Psi$ implies that  the only  non-vanishing terms are those with $\Gamma^a$ such that
$\Gamma_{ij}^a=-\Gamma_{ji}^a$: this relation holds for only six matrices ($a=3,7,9,10,12,15$). Furthermore, with the help of the Fierz identities we have
\beg\label{TaDam}
{\begin{split}
&\left[\begin{matrix}
\left({\Psi}\dg\Psi\right)^2\\ 
\left({\Psi}\dg \tau_1\sigma_0\Psi\right)^2 \\
\left({\Psi}\dg \tau_2\sigma_0\Psi\right)^2 \\
\left({\Psi}\dg \tau_3\sigma_0\Psi\right)^2 \\
\end{matrix}
\right]=
\frac{1}{4}\left(\begin{matrix}
1 & 1 & 1 & 1 & 1 & 1 \\
1 & 1 & -1 & -1 & -1 & -1 \\
-1 & 1 & -1 & -1 & -1 & 1 \\
1 & -1 & -1 & -1 & -1 & 1
\end{matrix}
\right)\\&\times
\left[\begin{matrix}
\left({\Psi}\dg(x)\Gamma^3{\Psi}^*(x)\right)\left({\Psi}^T(x)\Gamma^3\Psi(x)\right) \\
\left({\Psi}\dg(x)\Gamma^7{\Psi}^*(x)\right)\left({\Psi}^T(x)\Gamma^7\Psi(x)\right) \\
\left({\Psi}\dg(x)\Gamma^9{\Psi}^*(x)\right)\left({\Psi}^T(x)\Gamma^9\Psi(x)\right) \\
\left({\Psi}\dg(x)\Gamma^{10}{\Psi}^*(x)\right)\left({\Psi}^T(x)\Gamma^{10}\Psi(x)\right) \\
\left({\Psi}\dg(x)\Gamma^{12}{\Psi}^*(x)\right)\left({\Psi}^T(x)\Gamma^{12}\Psi(x)\right) \\
\left({\Psi}\dg(x)\Gamma^{15}{\Psi}^*(x)\right)\left({\Psi}^T(x)\Gamma^{15}\Psi(x)\right)
\end{matrix}
\right].
\end{split}}
\en
Introducing the following notations:
\beg\label{SingletTrip}\nonumber
\begin{split}
&{\cal S}_1(\br,\tau)=i{\Psi}^T(x)\Gamma^3\Psi(x)={\Psi}^T(x)(i\tau_0\sigma_2)\Psi(x), \\
&{\cal S}_2(\br,\tau)=i{\Psi}^T(x)\Gamma^{15}\Psi(x)={\Psi}^T(x)\left(i\tau_3\sigma_2\right)\Psi(x), \\
&{\cal T}_3(\br,\tau)=i{\Psi}^T(x)\Gamma^7\Psi(x)={\Psi}^T(x)\left(i\tau_1\sigma_2)\right)\Psi(x), \\
&{\cal T}_4^{(1)}(\br,\tau)=i{\Psi}^T(x)\Gamma^9\Psi(x)={\Psi}^T(x)\left(i\tau_2\sigma_0\right)\Psi(x), \\
& {\cal T}_4^{(2)}(\br,\tau)=i{\Psi}^T(x)\Gamma^{10}\Psi(x)={\Psi}^T(x)\left(i\tau_2\sigma_1\right)\Psi(x), \\
&{\cal T}_4^{(3)}(\br,\tau)=i{\Psi}^T(x)\Gamma^{12}\Psi(x)={\Psi}^T(x)\left(i\tau_2\sigma_3\right)\Psi(x).
\end{split}
\en
we can now write down the Lagrangian density describing the interactions in the particle-particle channel:
\beg\label{Lintpp}
{\begin{split}
{\cal L}_{\textrm{int}}&=\sum\limits_{j=1}^{2}{u}_j{\cal S}_j\dg(\br,\tau){\cal S}_j(\br,\tau)+
u_3{\cal T}_{3}\dg(\br,\tau){\cal T}_3(\br,\tau)\\&+
u_4\sum\limits_{m=1}^{3}{\cal T}_{4}^{(m)\dagger}(\br,\tau){\cal T}_4^{(m)}(\br,\tau).
\end{split}}
\en
where the newly introduced (pairing) coupling constants are
$u_1=(g_1+g_2-g_3+g_4)/4$, $u_2=(g_1-g_2+g_3+g_4)/4$,
$u_3=(g_1+g_2+g_3-g_4)/4$ and $u_4=(g_1-g_2-g_3-g_4)/4$.
Thus, just like in the case of particle-hole channel, we have ended up with four independent couplings. By expanding the operators (\ref{SingletTrip}) we can interpret ${\cal S}_1(\br,\tau)$ as the pairing operator in the $s$-wave channel, while ${\cal S}_2(\br,\tau)$ as the pairing operator leading to $s^{\pm}$-wave pairing. The remaining operators account for the pairing in either odd-parity and/or  spin-triplet channel. 

The renormalization group equations for the couplings $u_j$ can now be derived following the same procedure used to derive Eqs. (\ref{FlowEqs}).
It is worth mentioning here that in this case, that the first four diagram in Fig. \ref{FigDiags} give the same contribution (up to a numerical pre-factor) to the effective action and, importantly, only a contribution from the diagram (E) contains the mass anisotropy parameter $\eta$. The resulting RG equations in this case read:
\beg\label{Singlet}
\begin{split}
\frac{du_1}{d\ln s}&=\frac{m\Lambda}{2\pi^2}\left[(u_1-u_2)(u_3+3u_4)-2(1+2\eta)u_1^2\right], \\
\frac{du_2}{d\ln s}&=\frac{m\Lambda}{2\pi^2}\left[(u_2-u_1)(u_3+3u_4)-2(1+2\eta)u_2^2\right], \\
\frac{du_3}{d\ln s}&=\frac{m\Lambda}{4\pi^2}\left[(u_1-u_2)^2+u_3^2-3u_4^2+6u_3u_4\right], \\
\frac{du_4}{d\ln s}&=\frac{m\Lambda}{4\pi^2}\left[(u_1-u_2)^2+(u_3-u_4)^2+4u_4^2\right].
\end{split}
\en
As we have mentioned above these equations have been obtained independently of our earlier calculation, so one can readily check that upon expressing the coupling $u_j$'s in terms of the coupling constants $g_j$'s for the particle-hole channel interactions, we recover the RG equations  
(\ref{FlowEqs}).

\begin{figure}[h]
  \centering
 \includegraphics[width=3.15in]{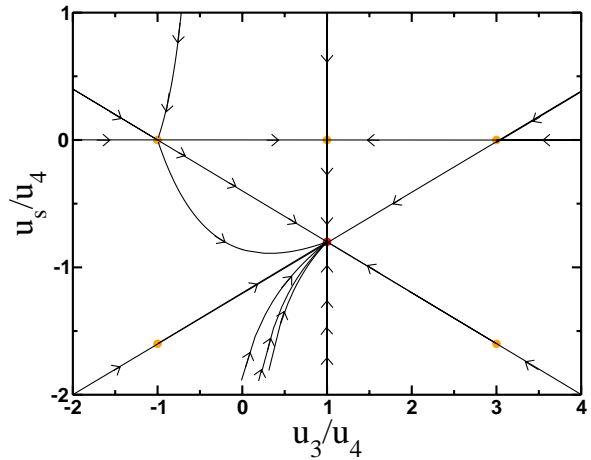}
  \caption{Renormalization group flow of the coupling constant ratios in the particle-particle channel for the case of small mismatch between the effective masses, $\eta=0.125$.
 We find that there are six fixed points overall in this case. Five fixed points (solid orange circles) are always unstable.  The remaning one (solid red circle) is stable when $u_4<0$ and becomes unstable when $u_4>0$. Without loss of generality we chose to limit the presentation to a case of $u_1=u_2=u_{s}$ and we also assumed that at the beginning of the RG flow $(m\Lambda/4\pi^2)u_4=-0.1$.}
  \label{Fig2}
\end{figure}

The fixed point(s) of the equations above (\ref{Singlet}) can be found using the same strategy as we have used above. Since the right-hand-side of the last equation (\ref{Singlet}) can be written as a sum of the squares, we consider the ratio of the couplings $\lambda_a=u_a/u_4$. We find that just like in the particle-hole case, there are six fixed points: five unstable ones and one stable ("sink") when initial value of the coupling $u_4<0$.
The results for the flow of the couplings are presented in Fig. \ref{Fig2}. 
The stable fixed point - "sink" - is determined by $\lambda_3^*=1$ and $\lambda_1^*=\lambda_2^*=(1/4)(2\lambda_3^*-(\lambda_3^*)^2-5)/(1+2\eta)$.

\subsection{Particle-particle channel susceptibilities}
To determine the leading channel for the pairing instability, we need to evaluate the corresponding susceptibilities. Introducing the source terms into the action
\beg\label{DeltaSpp}
\begin{split}
\Delta S_{\textrm{p-p}}&=-\sum\limits_{i}\Delta_{\Gamma^i}\int_k{\Psi}^T(\bk,\tau)\hat{\Gamma}^i\Psi(\bk,\tau),
\end{split}
\en
Here the summations is performed over matrices $i=3,7,9,10,12,15$ and $k=(\tau,\bk)$. The subsequent calculation is completely analogous to the one above for the susceptibilities in the particle-hole channel. Specifically, after integrating out the fast modes for the renormalization of the source term (\ref{DeltaS}) and keeping in mind that $\Gamma_{\nu\mu}^i=-\Gamma_{\mu\nu}^i$ we find
\beg\label{ReSource}
\begin{split}
&\Delta_{\Gamma^i}(s){\Psi}_<^T(k)\hat{\Gamma}\Psi_<(k)=s^2\Delta_{\Gamma^i}(1){\Psi}^T(k)\hat{\Gamma}^i\Psi(k)\\&-
s^2\Delta_{\Gamma^i}(1)\sum\limits_{\cal S}g_{\cal S}\int\frac{d\omega}{2\pi}\int\limits_{\Lambda/s}^\Lambda\frac{p^2dp}{2\pi^2}\\&\Gamma_{\nu\mu}^{(i)}G_{\mu\gamma}(\bp,i\omega){\cal S}_{\gamma\delta}\Psi_\delta(k)G_{\nu\alpha}(-\bp,-i\omega){\cal S}_{\alpha\beta}\Psi_\beta(k).
\end{split}
\en
The momentum and frequency integrals appearing here have been computed already (see Eqs. (\ref{pp},\ref{Cleq1}) in Appendix B). The equations
for the flow of the functions $\Delta_{\textrm{pp}}^{(i)}(s)$ are
\beg\label{FlowDeltaspp}
\begin{split}
&\frac{d\ln\Delta_{\Gamma^3}}{d\ln s}=2-(1+2\eta)(u_1+u_2+u_3-u_4)\frac{m\Lambda}{\pi^2}, \\
&\frac{d\ln\Delta_{\Gamma^{15}}}{d\ln s}=2-(1+2\eta)(u_1+u_2-u_3+u_4)\frac{m\Lambda}{\pi^2}, \\
&\frac{d\ln\Delta_{\Gamma^{a}}}{d\ln s}=2, ~(a=7,9,10,12).
\end{split}
\en
Since the only stable fixed point exists for $u_4<0$, the fastest divergent susceptibility is clearly determined by the ratio 
$(u_1+u_2+u_3-u_4)/(u_1+u_2-u_3+u_4)$.
Numerical analysis of these equations shows that susceptibility $\Delta_{\Gamma^3}$ corresponding to the singlet $s$-wave pairing is the one diverging fastest.
Furthermore, we find that while the leading divergence corresponds to the singlet pairing, the strongest divergence is still governed by the excitonic instability. 

\section{Conclusions}
As the recent experimental studies have shown, the materials which may exhibit the physical effects we have discussed so far are disordered either due to alloying or due to the presence of vacancies in the nominally stoichiometric compounds. This is especially relevant for the excitonic instabilities, which are prone to slightest anisotropy of the underlying band structure let alone the presence of disorder. Since our results so far ignored the presence of disorder, we cannot claim with certainty that the excitonic instability will still be the leading one in that case. This problem, however, requires a thoughtful and careful treatment and, as such, goes beyond the scope of the present study. 

Other avenues for further investigation of the problems related to the one discussed here concern the renormalization of the chemical potential especially when the system has been doped and, as a result, the superconducting instability develops faster than the excitonic one. Lastly, we would like to mention the situation when the $s$-orbital band inverts with the $f$-orbital one, which would mean the hybridization matrix element will have $V_\bk\propto k^3$, 
so upon integrating out the fast modes it will be the leading determining factor in renormalization of the coupling constants. With this being said, the specific focus of our the future studies depend mainly on the appearance of new experimental data. 

To conclude, we have studied the problem of weak coupling many-body instabilities in narrow gap semiconductors with odd-parity band inversion. Our study has been mainly motivated by recent experimental and theoretical work addressing thermodynamic properties of samarium hexaboride.
By employing the renormalization group technique we find that the leading instability is towards the formation of an excitonic order. Depending on the microscopic details of the model the leading excitonic instability may or may not break the time-reversal symmetry.

\section{Acknowledgments} We would like to thank Ammar Kirmani for initial collaboration on this project. Important discussions with Bitan Roy, Maxim Khodas and Oskar Vafek are gratefully acknowledged. MD expresses his gratitude to Max Planck Institute for Physics of Complex Systems (MPI-PKS, Dresden, Germany), where a part of this work has been completed, for hospitality. This work was financially supported by the National Science Foundation grant NSF-DMR-2002795 and, in part, by the U.S. Department of Energy, Basic Energy Sciences, grant DE-SC0016481.

\begin{appendix}
\section{Dirac matrices} 
We use the following definition of the Dirac matrices
\beg\label{DMat}
\begin{split}
\gamma_0&=\left(\begin{matrix} \hat{\sigma}_0 & 0 \\ 0 & -\hat{\sigma}_0\end{matrix}\right), \quad
\gamma_1=\left(\begin{matrix} 0 & \hat{\sigma}_x \\ - \hat{\sigma}_x & 0\end{matrix}\right), \\ 
\gamma_2&=\left(\begin{matrix} 0 &  \hat{\sigma}_y \\ -\hat{\sigma}_y & 0 \end{matrix}\right), \quad 
\gamma_3=\left(\begin{matrix} 0 &  \hat{\sigma}_z \\ - \hat{\sigma}_z & 0 \end{matrix}\right), \\
\gamma_5&=\left(\begin{matrix} 0 &  \hat{\sigma}_0 \\   \hat{\sigma}_0 & 0 \end{matrix}\right).
\end{split}
\en
Here $\hat{\sigma}_0$ is a 2$\times$2 unit matrix and $\hat{\sigma}_{a}$ ($a=x,y,z$) are Pauli matrices. 
\section{Renormalization group equations: auxiliary expressions}
\subsection{Cumulant expansion} 
To calculate the average entering into equation (\ref{Basic}) we will employ the cumulant expansion. It reads:
\beg\label{Cum}
{\langle e^{-S_{\textrm{int}}[\Psi_<,\Psi_>]}\rangle_{0>}\approx e^{-\langle S_{\textrm{int}}\rangle+\frac{1}{2}(\langle S_{\textrm{int}}^2\rangle-\langle S_{\textrm{int}}\rangle^2)+...}}
\en
To avoid the complications arising from the antisymmetrization of the interaction (\ref{LintRed}), we will formally consider the interaction part
of the action for general coupling in the form
\beg\label{Sintk}
\begin{split}
S_{\textrm{int}}&=\sum\limits_{{\cal S}{\cal T}}\prod\limits_{j=1,2}\int d\br_j\int d\tau_jU_{{\cal S}{\cal T}}(12)\\&\times\left({\Psi}\dg(1){\cal S}{\Psi}(1)\right)
\left(\Psi\dg(2){\cal T}\Psi(2)\right), 
\end{split}
\en
where we used the notation
\beg\label{UST}
\begin{split}
U_{{\cal S}{\cal T}}(12)&=\frac{g_{{\cal S}{\cal T}}}{2}\int d\tau\int d\br \prod\limits_{j=1}^2\delta(\br-\br_j)\delta(\tau-\tau_j)
\end{split}
\en
and we defined
$\Psi(j)=\Psi_{\alpha_j}(\br_j,\tau_j)$ ($j=1,2$).
Using these notations, for the correction to the action we find
\begin{widetext}
\beg\label{Action}\nonumber
\begin{split}
&\frac{1}{2}(\langle S_{\textrm{int}}^2\rangle-\langle S_{\textrm{int}}\rangle^2)=\frac{1}{2}
\sum\limits_{{\cal S}{\cal S}'}\sum\limits_{{\cal T}{\cal T}'}\sum\limits_{1234}U_{{\cal S}{\cal T}}(12)U_{{\cal S}'{\cal T}'}(34)
\langle\left({\Psi}\dg(1){\cal S}{\Psi}(1)\right)
\left(\Psi\dg(2){\cal T}\Psi(2)\right)\left({\Psi}\dg(3){\cal S}'{\Psi}(3)\right)
\left(\Psi\dg(4){\cal T}'\Psi(4)\right)\rangle\\&-
\frac{1}{2}\sum\limits_{{\cal S}{\cal S}'}\sum\limits_{{\cal T}{\cal T}'}\sum\limits_{1234}U_{{\cal S}{\cal T}}(12)U_{{\cal S}'{\cal T}'}(34)
\langle\left({\Psi}\dg(1){\cal S}{\Psi}(1)\right)
\left(\Psi\dg(2){\cal T}\Psi(2)\right)\rangle\langle\left({\Psi}\dg(3){\cal S}'{\Psi}(3)\right)
\left(\Psi\dg(4){\cal T}'\Psi(4)\right)\rangle
\end{split}
\en
\end{widetext}
and each $\Psi=\Psi_<+\Psi_>$. Note that the correction to the action is defined as
\beg\label{DefineSign}
{\Delta S_{\textrm{int}}=-\frac{1}{2}(\langle S_{\textrm{int}}^2\rangle-\langle S_{\textrm{int}}\rangle^2).}
\en
There are five different non-zero contributions to (\ref{Action}). The diagram describing the fist contribution is shown on Fig. \ref{FigDiags}.
\begin{figure}
\centering
\includegraphics[width=0.45\linewidth]{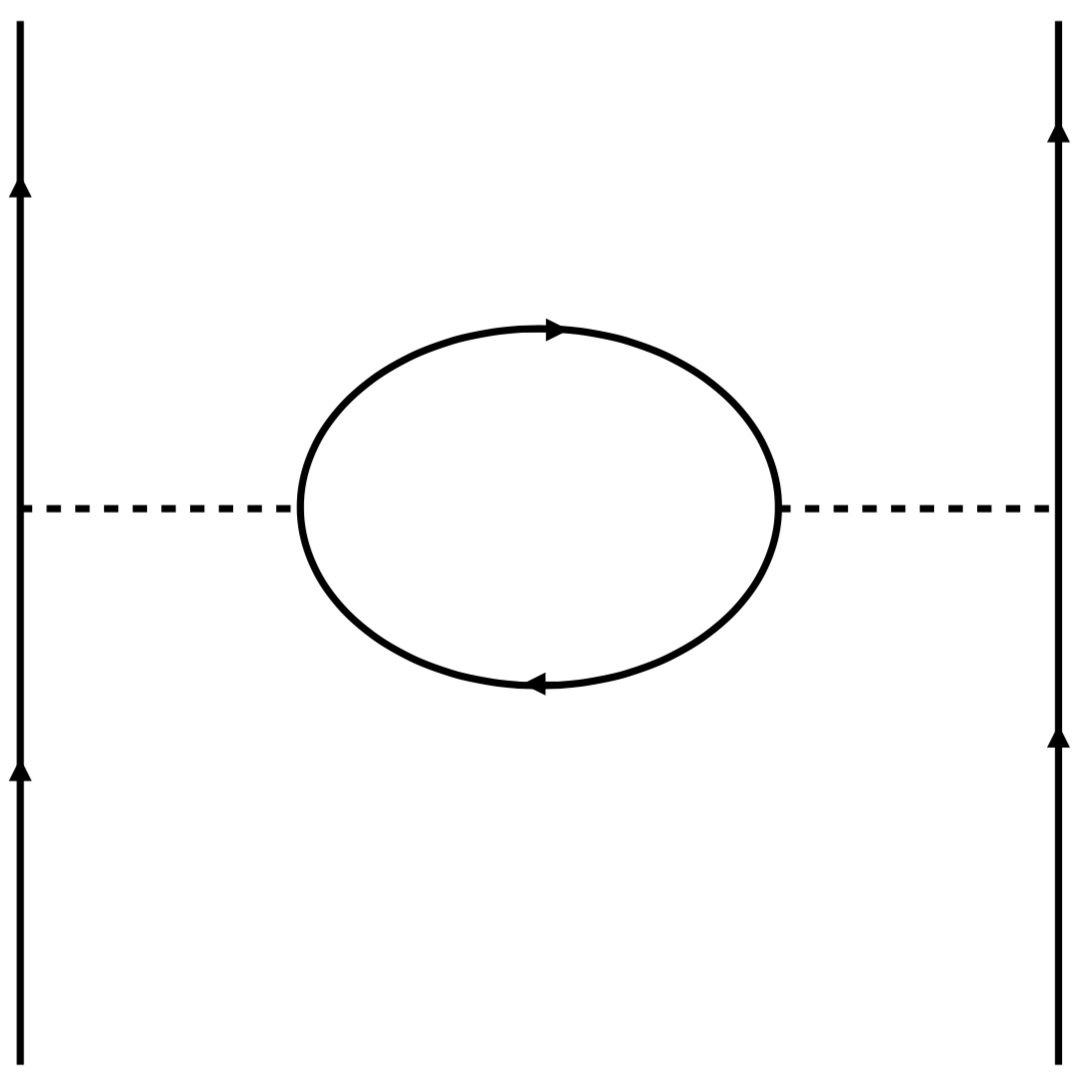}\\
\caption{Diagram containing a single fermionic loop, which appear in the expansion of the effective action (\ref{DefineSign}). The solid lines
represent the single-particle propagators, while the dashed lines represent the interaction (\ref{Sintk}). The momenta of the internal solid lines 
lie on the 'fast' momentum shell $\Lambda/s\leq k\leq\Lambda$.}
\label{FigDiagA}
\end{figure}
\begin{widetext}
\paragraph{Diagram A.} Analytical expression for the diagram Fig. \ref{FigDiagA} is given by 
\beg\label{DiagramA}
\begin{split}
&\frac{1}{2}\sum U_{{\cal S}_1{\cal T}_1}(12)U_{{\cal S}_2{\cal T}_2}(34)
\langle\left({\Psi}\dg(1){\cal S}_1{\Psi}(1)\right)
\left(\Psi\dg(2){\cal T}_1\Psi(2)\right)\left({\Psi}\dg(3){\cal S}_2{\Psi}(3)\right)
\left(\Psi\dg(4){\cal T}_2\Psi(4)\right)\rangle_A\\&=-\frac{1}{8}\sum\limits_{{\cal S}_1{\cal T}_1}\sum\limits_{{\cal S}_2{\cal T}_2}
g_{{\cal S}_1{\cal T}_1}g_{{\cal S}_2{\cal T}_2}\int_1\int_2\left\{\left({\Psi}\dg(1){\cal S}_1{\Psi}(1)\right)
\textrm{Tr}\left[{\cal T}_1G(1-2){\cal S}_2 G(2-1)\right]\left({\Psi}\dg(2){\cal T}_2{\Psi}(2)\right)
\right.\\&\left.+\left({\Psi}\dg(1){\cal S}_1{\Psi}(1)\right)
\textrm{Tr}\left[{\cal T}_1G(1-2){\cal T}_2 G(2-1)\right]\left({\Psi}\dg(2){\cal S}_2{\Psi}(2)\right)+
\left({\Psi}\dg(1){\cal T}_1{\Psi}(1)\right)
\right.\\&\left.\times\textrm{Tr}\left[{\cal S}_1G(1-2){\cal T}_2 G(2-1)\right]\left({\Psi}\dg(2){\cal S}_2{\Psi}(2)\right)+
\left({\Psi}\dg(1){\cal T}_1{\Psi}(1)\right)
\textrm{Tr}\left[{\cal S}_1G(1-2){\cal S}_2 G(2-1)\right]\left({\Psi}\dg(2){\cal T}_2{\Psi}(2)\right)
\right\}\\&=-\frac{1}{2}\sum\limits_{{\cal S}_1{\cal S}_2}g_{{\cal S}_1}g_{{\cal S}_2}\int_1\int_2\left({\Psi}\dg(1){\cal S}_1{\Psi}(1)\right)
\textrm{Tr}\left[{\cal S}_1G(1-2){\cal S}_2 G(2-1)\right]\left({\Psi}\dg(2){\cal S}_2{\Psi}(2)\right).
\end{split}
\en
\paragraph{Diagrams B \& C.} The correction to the action from the diagram ${\bf (B)}$ , Fig. \ref{FigDiags}(b,c), reads:
\beg\label{DiagramB}
\begin{split}
&\frac{1}{2}\sum U_{{\cal S}_1{\cal T}_1}(12)U_{{\cal S}_2{\cal T}_2}(34)
\langle\left({\Psi}\dg(1){\cal S}_1{\Psi}(1)\right)
\left(\Psi\dg(2){\cal T}_1\Psi(2)\right)\left({\Psi}\dg(3){\cal S}_2{\Psi}(3)\right)
\left(\Psi\dg(4){\cal T}_2\Psi(4)\right)\rangle_B\\&=\frac{1}{8}\sum\limits_{{\cal S}_1{\cal T}_1}\sum\limits_{{\cal S}_2{\cal T}_2}
g_{{\cal S}_1{\cal T}_1}g_{{\cal S}_2{\cal T}_2}\int_1\int_2
\times\left\{\left({\Psi}\dg(1){\cal S}_1G(1-2){\cal S}_2 G(2-1){\cal T}_1{\Psi}(1)\right)\left({\Psi}\dg(2){\cal T}_2{\Psi}(2)\right)\right.\\&\left.+\left({\Psi}\dg(1){\cal T}_1G(1-2){\cal S}_2 G(2-1){\cal S}_1{\Psi}(1)\right)\left({\Psi}\dg(2){\cal T}_2{\Psi}(2)\right)+
\left({\cal S}_2\leftrightarrow{\cal T}_2\right)\right\}\\&=\frac{1}{2}\sum\limits_{{\cal S}_1{\cal S}_2}g_{{\cal S}_1}g_{{\cal S}_2}\int_1\int_2
\left({\Psi}\dg(1){\cal S}_1G(1-2){\cal S}_2 G(2-1){\cal S}_1{\Psi}(1)\right)\left({\Psi}\dg(2){\cal S}_2{\Psi}(2)\right).
\end{split}
\en
Finally, the last two contributions to the action can be described by the two diagrams in Fig. \ref{FigDiags}(d,e). 
For the diagram ${\bf (D)}$ we derive the following expression
\beg\label{DiagramD3}
\begin{split}
&\frac{1}{2}\sum U_{{\cal S}_1{\cal T}_1}(12)U_{{\cal S}_2{\cal T}_2}(34)
\langle\left({\Psi}\dg(1){\cal S}_1{\Psi}(1)\right)
\left(\Psi\dg(2){\cal T}_1\Psi(2)\right)\left({\Psi}\dg(3){\cal S}_2{\Psi}(3)\right)
\left(\Psi\dg(4){\cal T}_2\Psi(4)\right)\rangle_D\\&=\frac{1}{2}\sum\limits_{{\cal S}_1{\cal S}_2}g_{{\cal S}_1}g_{{\cal S}_2}\int_1\int_2
\left(\Psi\dg(1){\cal S}_1G(1-2){\cal S}_2\Psi(2)\right)\left(\Psi\dg(2){\cal S}_2G(2-1){\cal S}_1\Psi(1)\right).
\end{split}
\en

Similarly, for the diagram ${\bf (E)}$  we find
\beg\label{DiagramE}
\begin{split}
&\frac{1}{2}\sum U_{{\cal S}_1{\cal T}_1}(12)U_{{\cal S}_2{\cal T}_2}(34)
\langle\left({\Psi}\dg(1){\cal S}_1{\Psi}(1)\right)
\left(\Psi\dg(2){\cal T}_1\Psi(2)\right)\left({\Psi}\dg(3){\cal S}_2{\Psi}(3)\right)
\left(\Psi\dg(4){\cal T}_2\Psi(4)\right)\rangle_E\\&=\frac{1}{4}\sum\limits_{{\cal S}_1{\cal S}_2}g_{{\cal S}_1}g_{{\cal S}_2}\int_1\int_2
\left\{\left(\Psi\dg(1){\cal S}_1G(1-2){\cal S}_2\Psi(2)\right)^2+\left(\Psi\dg(1){\cal S}_2G(1-2){\cal S}_1\Psi(2)\right)^2\right\}.
\end{split}
\en
We would like to remind the reader that the integration in the internal fermionic lines is limited to the momentum shell $[\Lambda/s,\Lambda]$.
\begin{figure}
\centering
\includegraphics[width=0.45\linewidth]{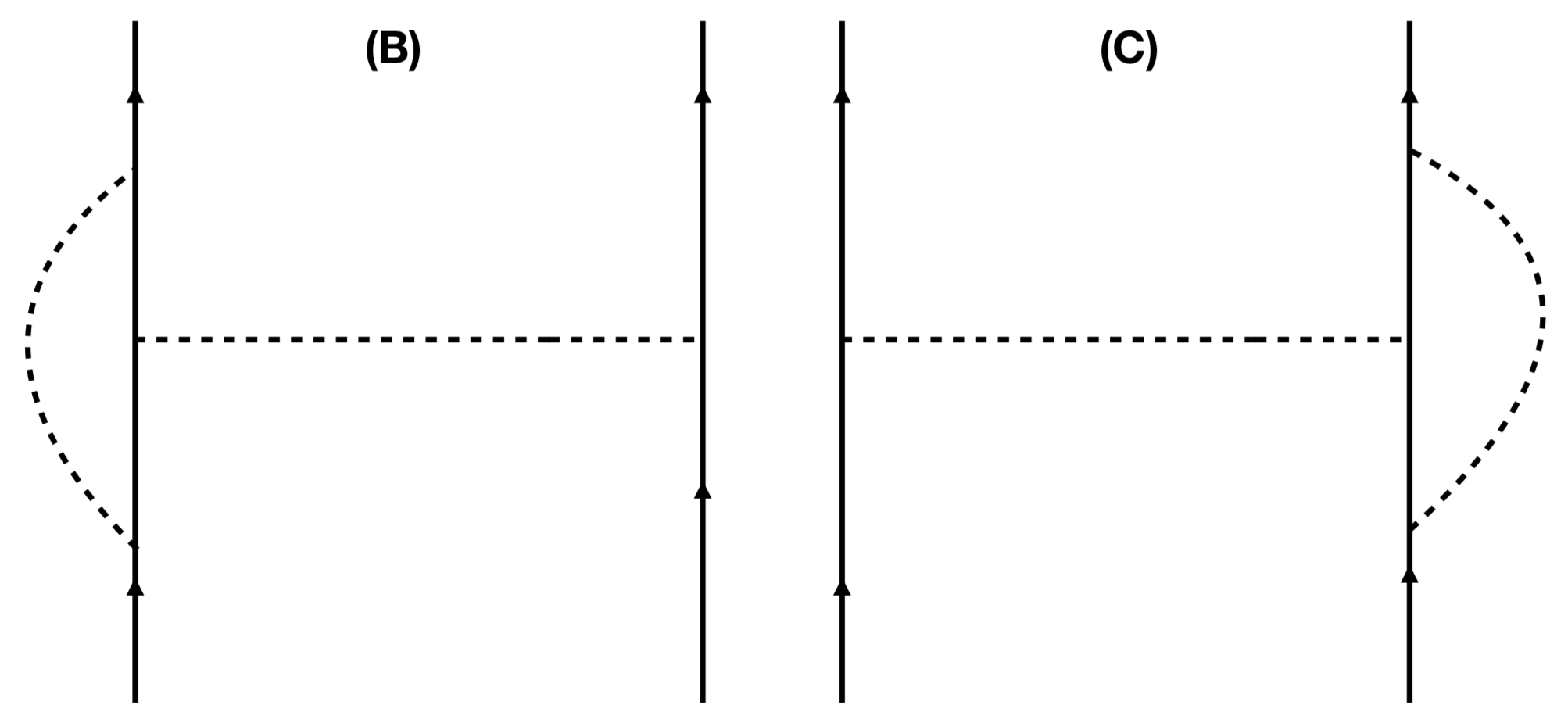}
~\includegraphics[width=0.38\linewidth]{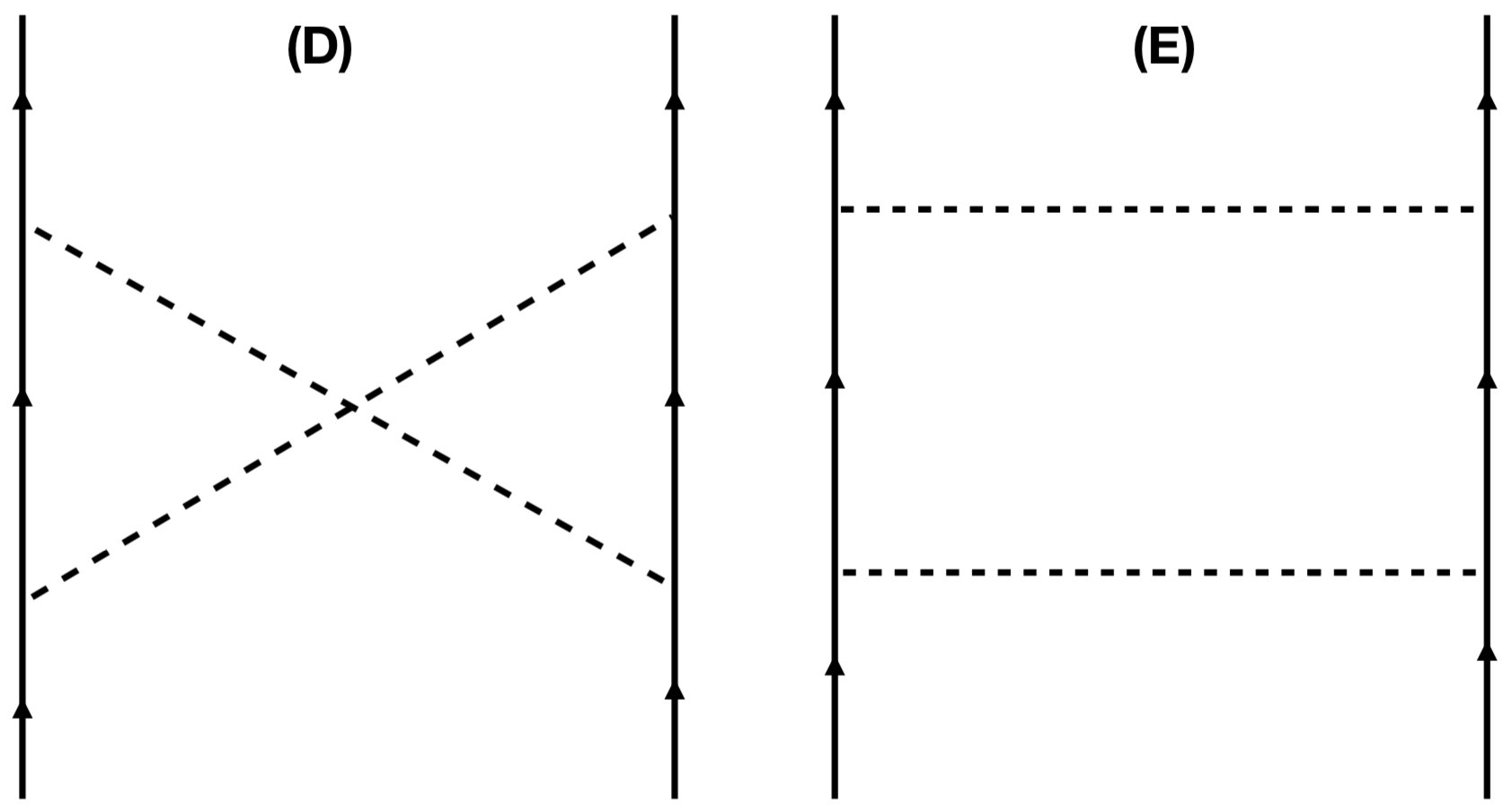}
\caption{Remaining four diagrams in one-loop approximation, which appear in the expansion of the effective action (\ref{DefineSign}). The solid lines
represent the single-particle propagators, while the dashed lines represent the interaction (\ref{Sintk}). The momenta of the internal solid lines 
lie on the 'fast' momentum shell $\Lambda/s\leq k\leq\Lambda$.}
\label{FigDiags}
\end{figure}

\end{widetext}
\subsection{Single-particle propagator}
These expressions can now be used integrate out the fast modes. To do that, we use the expression for the single particle propagator
\beg\label{Gkw}
\begin{split}
&G_\bk(i\omega_n)=\left((i\omega_n+\mu-\xi_\bk)\mathbbm{1}_4-\sum\limits_{a}\Sigma^ad_\bk^a\right)^{-1}\\&=-\frac{(i\omega_n+\mu-\xi_\bk)\mathbbm{1}_4+\gamma_0d_\bk^0+\sum\limits_a\gamma_0\gamma_ad_\bk^a}{(\omega_n-i\mu+i\xi_\bk)^2+E_\bk^2},
\end{split}
\en
where $\omega_n=\pi T(2n+1)$ are Matsubara frequencies and
${E_\bk=\sqrt{(d_\bk^0)^2+(d_\bk^x)^2+(d_\bk^y)^2+(d_\bk^z)^2}}$
is the renormalized single-particle spectrum. 
\subsection{Particle-hole channel at $T=0$}
Here we will evaluate the one-loop diagrams on the momentum shell $p\in[\Lambda/s,\Lambda]$. 
Recall that in the limit $T\to0$
\beg\label{sum2int}
T\sum\limits_{\omega_n} \to \int\frac{d\omega}{2\pi} 
\en
We adopted the following notations $s=e^t$, so for the infinitesimally narrow shell $\Lambda/s=\Lambda e^{-\delta t}\approx \Lambda(1-\delta t)$.
Next we consider an expression for the fermionic loop in particle-hole channel
\beg\label{Pph}
\hat{\cal P}_l(\Lambda,\mu)=\int\limits_{\Lambda(1-\delta t)}^{\Lambda}\frac{k^2dk}{2\pi^2}\int\frac{d\Omega_\bk}{4\pi}\int\limits_{-\infty}^{\infty}\frac{d\omega}{2\pi}G_\bk(i\omega)\otimes G_\bk(i\omega)
\en
and here we use the compact notation $G\otimes G\equiv G_{\alpha_1\alpha_2}G_{\alpha_3\alpha_4}$.
Integration over frequency
\beg\label{omegaint}
\begin{split}
&\int\limits_{-\infty}^{\infty}\frac{d\omega_n}{2\pi}\frac{(i\omega_n+\mu-\xi_\bk)^2}{[(\omega_n+i(\xi_\bk-\mu)^2+E_\bk^2]^2}\\&
=
\frac{1}{4E_\bk^3}\left[\vartheta(x_1+1)-\vartheta(x_1-1)\right],
\end{split}
\en
where $x_1=(\mu-\xi_\bk)/E_\bk$.
It will also be convenient to work with function $F_1(x)$ which is defined according to:
\beg\label{defF0}
{F_1(x)=\frac{1}{2}\textrm{sign}(1+x_1)+\frac{1}{2}\textrm{sign}(1-x_1).}
\en

It is straightforward to integrate over frequency which yields ($\delta t\ll 1$):
\beg\label{ph}
\begin{split}
&\hat{\cal P}_1(\Lambda,\mu)=\int\limits_{\Lambda(1-\delta t)}^{\Lambda}\frac{k^2dk}{2\pi^2}\int\frac{d\Omega_\bk}{4\pi}F_1\left(\frac{\mu-\xi_\bk}{E_\bk}\right)\times\\&\left\{-\frac{1}{4E_\bk}\left(\mathbbm{1}_4\otimes\mathbbm{1}_4\right)+\frac{1}{4E_\bk^3}\sum\limits_{ab}d_\bk^a d_\bk^b\left(\Sigma^a\otimes\Sigma^b\right)\right\}.
\end{split}
\en
To the leading order in powers of $\Lambda\gg k_F$ the hybridization term is much smaller than the kinetic energy:
\beg\label{ph2}
\begin{split}
&\hat{\cal P}_1(\Lambda,\mu)=\int\limits_{\Lambda(1-\delta t)}^{\Lambda}\frac{k^2dk}{8\pi^2}F_1\left(\frac{\mu-\xi_\bk}{E_\bk}\right)\\&\times\left\{
\frac{\eps_\bk^2}{E_\bk^3}\left(\Sigma^0\otimes\Sigma^0\right)
-\frac{\mathbbm{1}_4\otimes \mathbbm{1}_4}{E_\bk}\right\}.
\end{split}
\en 
The value of the remaining integral can be estimated by taking $\Lambda\to\infty$ and $\delta t\ll 1$. I have
\beg\label{PT0}
\hat{\cal P}_{1}(\Lambda,\mu)\approx\frac{m_{+}\Lambda}{4\pi^2}F_1\left(-\frac{m_{+}}{m_{-}}\right)
\left\{\gamma_0\otimes\gamma_0-\mathbbm{1}_4\otimes \mathbbm{1}_4\right\}\delta t.
\en
Since $m_{+}/m_{-}=(m_f-m_c)/(m_f+m_c)$, in the limit $m_f\gg m_c$ it follows that $m_{+}/m_{-}\approx 1-2m_c/m_f$, so that $F_1(-m_{+}/m_{-})\approx 1$. 
Note that the pre-factor  is proportional to the density of states at the Fermi level per spin for the free electrons in three-dimensions. 

\subsection{Particle-particle channel at $T=0$}
For the computation of the diagrams ${\bf (D)}$ and ${\bf (E)}$ I will also need to compute the following integral ($\delta t\ll 1$):
\beg\label{pp}
\begin{split}
&\hat{K}_1(\Lambda,\mu)=\int\limits_{\Lambda(1-\delta t)}^{\Lambda}\frac{k^2dk}{2\pi^2}\int\frac{d\Omega_\bk}{4\pi}\\&\times\int\limits_{-\infty}^{\infty}\frac{d\omega}{2\pi}G_\bk(i\omega)\otimes G_{-\bk}(-i\omega)
\end{split}
\en
Just like for the calculation of the particle-hole loop, we will integrate over $\omega$ first and write down the 
results in terms of the following functions:
\beg\label{AuxInts}
\begin{split}
&\int\limits_{-\infty}^\infty\frac{d\omega}{2\pi}\frac{1}{[(\omega+i\mu-i\xi_\bk)^2+E_\bk^2][(\omega-i\mu+i\xi_\bk)^2+E_\bk^2]}\\&
=\frac{{\cal C}_1^{(1)}[(\mu-\xi_\bk)/E_\bk]}{4E_\bk^3}, \\
&\int\limits_{-\infty}^\infty\frac{d\omega}{2\pi}\frac{[\omega^2+(\mu-\xi_\bk)^2]}{[(\omega+i\mu-i\xi_\bk)^2+E_\bk^2][(\omega+i\xi_\bk-i\mu)^2+E_\bk^2]}\\&=\frac{{\cal C}_1^{(2)}[(\mu-\xi_\bk)/E_\bk]}{4E_\bk}.
\end{split}
\en
where functions ${\cal C}_1^{(1)}$ and ${\cal C}_1^{(2)}$ are defined by
\beg\label{defF1F2}
\begin{split}
&{\cal C}_1^{(1)}(x)=\frac{x\vartheta(1-x)-\vartheta(x-1)}{x(1-x^2)}, \\ 
&{\cal C}_1^{(2)}(x)=\frac{x\vartheta(1-x)+(1-2x^2)\vartheta(x-1)}{x(1-x^2)}.
\end{split}
\en
We find
\beg\label{Ipp}
\begin{split}
&\hat{K}_1(\Lambda,\mu)=\int\limits_{\Lambda(1-\delta t)}^{\Lambda}\frac{k^2dk}{2\pi^2}
\left\{\frac{\eps_\bk^2}{4E_\bk^3}\left(\Sigma^0\otimes\Sigma^0\right){\cal C}_1^{(1)}\left(\frac{\mu-\xi_\bk}{E_\bk}\right)
\right.\\&\left.+\frac{1}{4E_\bk}\left(\mathbbm{1}_4\otimes \mathbbm{1}_4\right){\cal C}_1^{(2)}\left(\frac{\mu-\xi_\bk}{E_\bk}\right)\right\}.
\end{split}
\en
Just like in our analysis of the particle-hole channel, by taking into consideration $\Lambda^2/2m\mu\gg 1$ and $\delta t\ll 1$, we arrive to the following expression
\beg\label{Cleq1}
\begin{split}
&\hat{K}_{1}(\Lambda,\mu)\approx\frac{m_{+}\Lambda}{4\pi^2}\left\{\left(\gamma_0\otimes\gamma_0\right){\cal C}_1^{(1)}\left(-\frac{m_{+}}{m_{-}}\right)\right.\\&\left.+\left(\mathbbm{1}_4\otimes \mathbbm{1}_4\right){\cal C}_1^{(2)}\left(-\frac{m_{+}}{m_{-}}\right)\right\}\delta t+O(\delta t^2).
\end{split}
\en
This expression can be further simplified since is usually $m_f\gg m_c$ so that:
\beg\label{masses}
\begin{split}
\frac{m_{+}}{m_{-}}&=\frac{m_fm_c}{(m_f+m_c)}\frac{(m_f-m_c)}{m_fm_c}=\frac{m_f-m_c}{m_f+m_c} \leq 1.
\end{split}
\en
Then it follows
\beg\label{C1C2simp}
\begin{split}
&{\cal C}_1^{(1)}\left(-\frac{m_{+}}{m_{-}}\right)\\&={\cal C}_1^{(2)}\left(-\frac{m_{+}}{m_{-}}\right)\\&=\frac{(m_f+m_c)^2}{4m_fm_c}\approx\frac{m_f}{4m_c}\equiv{1+2\eta}.
\end{split}
\en
We use these results to compute the corrections to the coupling constants. 
\end{appendix}
\bibliography{rgl1bib}
\end{document}